\documentclass[]{spie}  %>>> use for US letter paper

\usepackage{amsmath,amsfonts,amssymb}
\usepackage{graphicx}
\usepackage[colorlinks=true, allcolors=blue]{hyperref}

\title{Lessons for WFIRST CGI from ground-based high-contrast systems}

\author[1]{Vanessa P. Bailey}
\author[1]{Michael Bottom}
\author[1]{Eric Cady}
\author[2]{Faustine Cantalloube}
\author[3]{Jozua de Boer}
\author[4]{Tyler Groff}
\author[1]{John Krist}
\author[1, a]{Maxwell A. Millar-Blanchaer}
\author[5]{Arthur Vigan}
\author[6, 7]{Jeffrey Chilcote}
\author[8, a]{Elodie Choquet}
\author[9]{Robert J. De Rosa}
\author[10, 11]{Julien H Girard}
\author[12]{Olivier Guyon}
\author[1]{Brian Kern}
\author[11]{Anne-Marie Lagrange}
\author[7]{Bruce Macintosh}
\author[13]{Jared R. Males}
\author[14]{Christian Marois}
\author[8, 15]{Tiffany Meshkat}
\author[16]{Julien Milli}
\author[17]{Mamadou N'Diaye}
\author[14]{Henry Ngo}
\author[7]{Eric L. Nielsen}
\author[1]{Jason Rhodes}
\author[8, b]{Garreth Ruane}
\author[3]{Rob G. van Holstein}
\author[9]{Jason J. Wang}
\author[8]{Wenhao Jerry Xuan}

\affil[1]{Jet Propulsion Laboratory, California Institute of Technology}
\affil[2]{Max Planck Institute for Astronomy}
\affil[3]{Leiden Observatory}
\affil[4]{NASA Goddard Space Flight Center}
\affil[5]{Aix Marseille Univ., CNRS, CNES, LAM, Marseille, France}
\affil[6]{University of Notre Dame}
\affil[7]{Kavli Institute of Particle Astrophysics and Cosmology, Stanford University}
\affil[8]{California Institute of Technology}
\affil[9]{University of California Berkeley}
\affil[10]{Space Telescope Science Institute}
\affil[11]{Universit\'e Grenoble Alpes, CNRS, IPAG}
\affil[12]{Subaru Telescope}
\affil[13]{University of Arizona, Steward Observatory}
\affil[14]{National Research Council of Canada}
\affil[15]{Infrared Processing and Analysis Center}
\affil[16]{European Southern Observatory}
\affil[17]{Universit\'e C\^ ote d'Azur, Observatoire de la C\^ ote d'Azur, CNRS, Laboratoire Lagrange}
\affil[a]{NASA Hubble Fellow}
\affil[b]{NSF Fellow}

\authorinfo{For further information, send correspondence to Vanessa Bailey: vanessa.bailey@jpl.nasa.gov}

\begin{document} 
\maketitle

\begin{abstract}
The Coronagraph Instrument (CGI) for NASA's Wide Field Infrared Survey Telescope (WFIRST) will constitute a dramatic step forward for high-contrast imaging, integral field spectroscopy, and polarimetry of exoplanets and circumstellar disks, aiming to improve upon the sensitivity of current ground-based direct imaging facilities by 2--3 orders of magnitude. Furthermore, CGI will serve as a pathfinder for future exo-Earth imaging and characterization missions by demonstrating wavefront control, coronagraphy, and spectral retrieval in a new contrast regime, and by validating instrument and telescope models at unprecedented levels of precision. To achieve this jump in performance, it is critical to draw on the experience of ground-based high-contrast facilities. We discuss several areas of relevant commonalities, including: wavefront control, post-processing of integral field unit data, and calibration and observing strategies.
\end{abstract}

\section{INTRODUCTION}
\label{sec:intro}  

\begin{figure} [ht]
   \begin{center}
   \begin{tabular}{c} %% tabular useful for creating an array of images 
   \includegraphics[height=9cm]{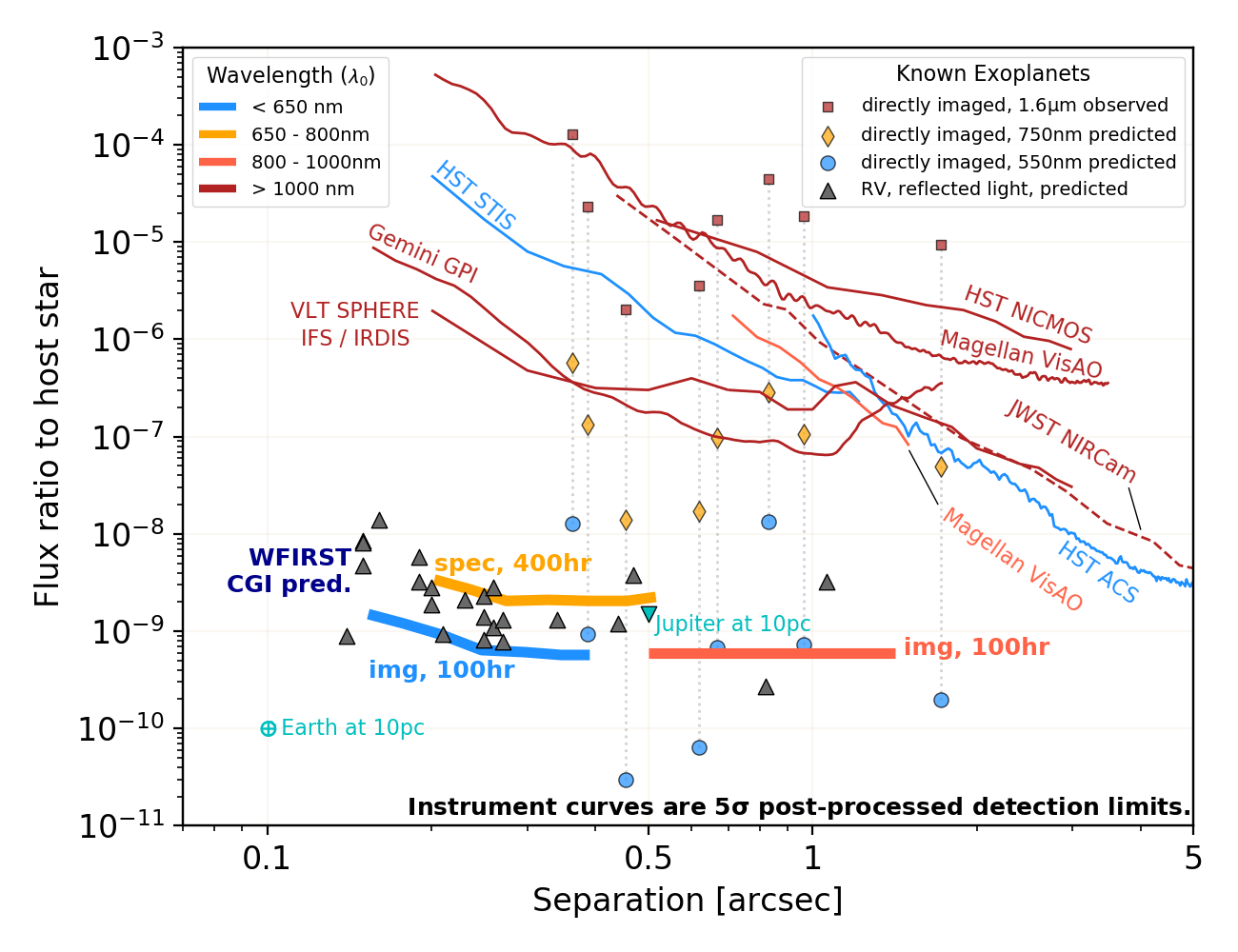}
   \end{tabular}
   \end{center}
   \caption[example] { \label{fig:theplot} 
Predicted CGI performance on a V=5 star, in the context of current high-contrast instrumentation\cite{ThePlot}. Curves and planets are color-coded by bandpass central wavelength. Curves are $5\sigma$ post-processed point source detection limits; bold lines are predictions for the three official CGI observing modes. CGI integration times are noted in the plot, while other instruments' performances are typically based on $\sim1$~hr of integration time. Known self-luminous imaged planets (colored points) are shown at their observed IR flux ratios as well as at their predicted flux ratios at visible wavelengths.  Gray triangles are predicted flux ratios for known giant planets detected by the radial velocity technique, when viewed at quadrature with assumed albedoes of 0.5.}
   \end{figure}

The Wide Field Infrared Survey Telescope (WFIRST) Coronagraph Instrument (CGI) is a high-contrast imager, polarimeter, and integral field spectrograph (IFS) that will enable the study of exoplanets and circumstellar disks at visible wavelengths ($\sim 550-850$~nm).\footnote{Instrument parameters and simulations are available at \url{https://wfirst.ipac.caltech.edu/}} 
CGI aims to achieve detection limits of $\sim 10^{-9}$ the flux of the host star at separations of $\sim 0.15''-1.5''$ (Figure \ref{fig:theplot}), in order to serve as a pathfinder for future terrestrial planet finding missions that require $\sim 10^{-10}$ flux ratio capability. If CGI achieves its predicted performance, it will be capable of imaging and low-resolution spectroscopy of gas giant planets in reflected light and of detection of exozodiacal dust disks at unprecedented sensitivity. These observations would begin to constrain cloud properties of mature Jupiter analogues, to shed light on the planet formation process in protoplanetary disks, and to identify the ``cleanest'' (least dusty) systems for future exo-Earth searches. To achieve these goals, WFIRST CGI will build not only on a legacy of previous space observatories, but also on a legacy of ground-based high-contrast instrumentation.  We refer the reader to Debes et al., 2015 \cite{Debes2015} for an excellent discussion of lessons learned from the Hubble and James Webb Space Telescopes. In these proceedings, we discuss relevant lessons learned from ground-based instruments in the areas of wavefront sensing and control, observing strategies, calibration, and data post-processing.

\section{WAVEFRONT SENSING AND CONTROL}
\label{sec:wavefront control}

At small working angles, contrast is governed by star centering behind the coronagraph occulting spot and by low order aberrations. 
Control of both fast tip/tilt jitter and of slower drift in low order modes is necessary in order to achieve the best performance. 
Several ground-based instruments have dedicated low order wavefront sensing and control (LOWFS/C) systems, including P1640\cite{Cady2013}, SPHERE\cite{NDiaye2016}, SCExAO\cite{Singh2015}, and GPI\cite{Hartung2014}. 
One common lesson learned from all of these systems is that for optimal performance, LOWFS/C must be operated continuously, in parallel with science observations. 
The LOWFS should not take any light from the science camera, and so designs utilizing light rejected by the coronagraph are preferred.
Furthermore, the LOWFS/C loop speed must be matched to the power spectrum of the appropriate modes (both fast tip/tilt jitter and slower drifts in higher modes).
In CGI, tip/tilt errors will originate from both sub-Hz observatory pointing drift and from structural vibrations excited by the telescope reaction wheels (1-100Hz). 
Longer timescale thermal drifts in the spacecraft will be the primary contributors to errors in other low order modes.
To compensate, CGI will have a dedicated LOWFS/C system for Zernike modes 2-11\cite {Shi2016}. 
The Zernike phase contrast wavefront sensor will use the starlight reflected by the coronagraph occulting masks and will operate during science observations. 
A fast steering mirror will correct tip/tilt jitter at frequencies $\lesssim 20$~Hz; other modes will be corrected at 5~mHz with a combination of a dedicated focus corrector and deformable mirrors (DMs). 

\begin{figure} [ht]
   \begin{center}
   \begin{tabular}{c} %% tabular useful for creating an array of images 
   \includegraphics[width=0.7\textwidth]{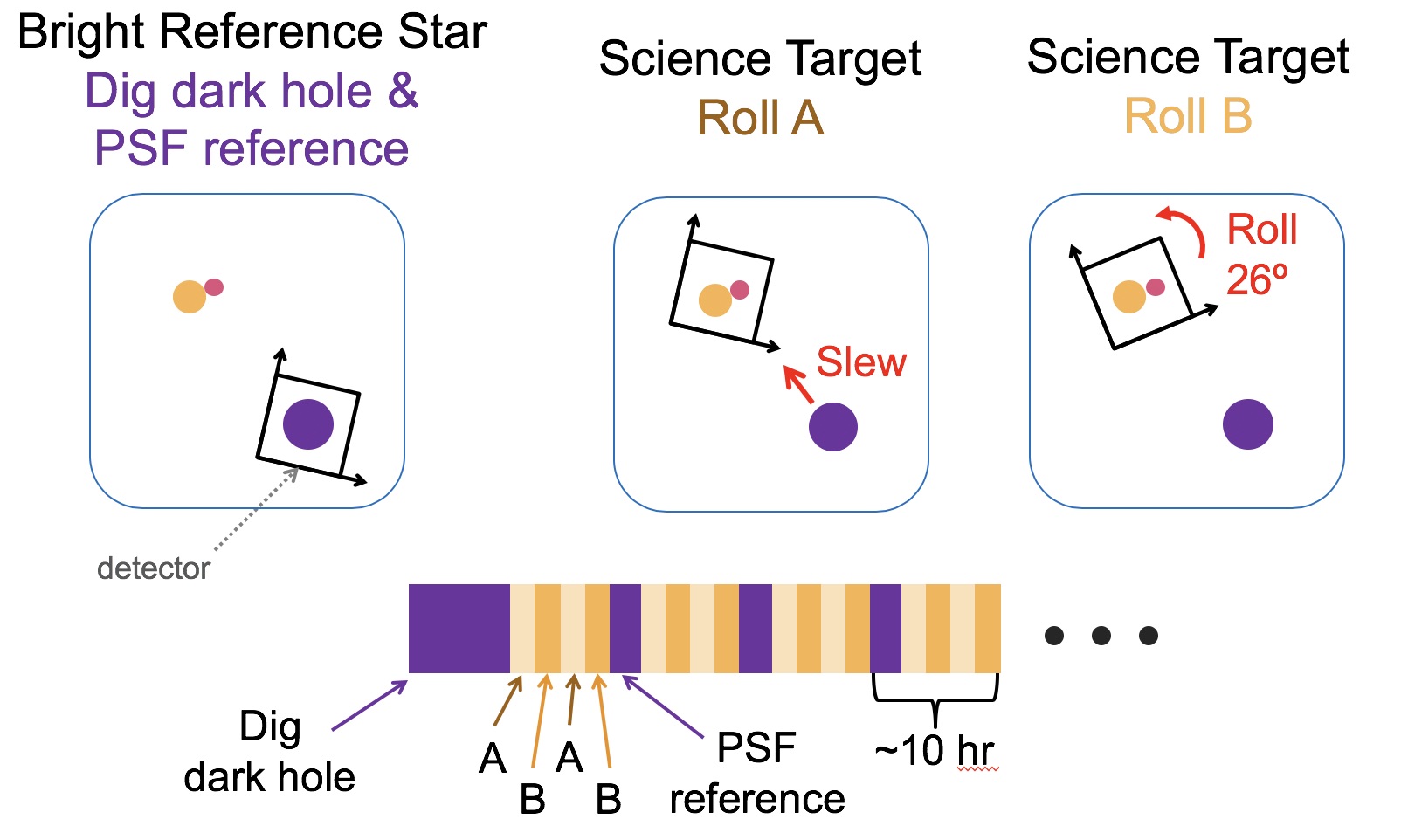}
   \end{tabular}
   \end{center}
   \caption[example]{ \label{fig:os6} CGI baseline observing scenario. CGI will alternate between one or more bright reference stars and the science target, typically separated by $15^\circ - 20^\circ$ on the sky. The initial dark hole will be optimized on the first visit to the reference star. The high-order wavefront correction will be frozen throughout the science target observation, and will be re-optimized on each subsequent reference star visit. CGI will alternate between two roll angles on the science target. Simulated images and IFS data for this observing scenario are \href{https://wfirst.ipac.caltech.edu/sims/Coronagraph_public_images.html}{publicly available.}}
\end{figure}

At larger working angles in the so-called ``dark hole,'' a High Order Wavefront Sensing and Control (HOWFS/C) system is needed.
All ground-based high-contrast instruments have dedicated high order wavefront sensors to sense atmospheric turbulence; CGI does not have a need for an analogous system.
CGI does, however, require HOWFS/C to counteract optical aberrations in the telescope and instrument\footnote{In ground-based systems, such internal aberrations are often referred to as ``non-common path aberrations,'' because they are not in the optical path of the Adaptive Optics system's dedicated wavefront sensor.}.
These aberrations are best sensed using the images from the science camera itself.
Several focal plane wavefront sensing techniques have been tested on P1640, Keck, SCExAO, and GPI\cite{Martinache2014, Matthews2017, Savransky2012a, Bottom2016a, Bottom2017}.
However, they have achieved varying degrees of success on sky, because precise knowledge of the instrument model and deformable mirror calibration is required.
Furthermore, drifts in the instrument cause evolution of the point spread function (PSF), and so the quality of the dark hole degrades without continuous HOWFS/C.
CGI has elected to use pairwise probing and electric field conjugation; instrument model calibration is the subject of ongoing efforts\cite{Giveon2011a, Seo2017, Cady2017}.
CGI will be photon-starved on science targets; hence, the current baseline is to dig the dark hole on a bright reference star and to freeze the high-order DM correction throughout the science sequence (Figure \ref{fig:os6}). 
Ongoing work is assessing the requirements on instrument thermal stability and on DM calibration for this ``optimize and freeze'' operational scenario.

\section{OBSERVING STRATEGY AND PSF SUBTRACTION}

For any high-contrast imager, the observing and post-processing strategies are interdependent. 
To recover faint sources, one must first subtract the PSF of the primary star.
Several different strategies for PSF synthesis and subtraction have been developed on ground-based instruments, and each has a corresponding observing strategy.
The differential imaging techniques are then combined with PSF synthesis algorithms such as a median combination of images, LOCI\cite{Lafreniere2007}, or KLIP\cite{Soummer2007}.

\subsection{Angular Differential Imaging}
In ADI\cite{Marois2006}, the telescope pupil rotates with respect to the sky. The stellar PSF is fixed in detector coordinates, while planets or disks appear rotate around the central star. 
A minimum amount of rotation, $\theta$, is required to avoid self-subtraction of planet light, typically $r\theta \sim 1$~FWHM (Full Width at Half Max), where $r$ is the star-planet separation. 
This observing strategy is common practice for high-contrast ground-based instruments. 
CGI can roll a maximum of $26^\circ$ due to Sun-to-spacecraft angle constraints, limiting the effectiveness of ADI at small working angles.
Unlike ground-based instruments, CGI will not have continuous roll angle coverage.
Instead, it will roll between the two extreme possible angles\footnote{Ongoing work is investigating the utility of adding a small number of intermediate roll angles.} (Figure \ref{fig:os6}), in a strategy more typical of space-based observatories.
The PSF of the coronagraph used in IFS mode has extended sidelobes in the azimuthal direction, so CGI's limited roll angle means that ADI cannot be used near the inner edge of the field of view of the IFS. 
Additionally, regardless of the coronagraph design, ADI acts as a high-pass filter on extended sources, and so it is not optimal for disks. 
For these reasons, CGI must employ Reference Differential Imaging as well.

\subsection{Reference Differential Imaging}
An RDI observing sequence alternates between the science target and one or more reference targets, and the reference target images are used for PSF synthesis.
RDI can be used in concert with ADI (RDI+ADI) if the science target images are included in the reference library used for PSF synthesis. 
Even in pure RDI, observing the science target at multiple roll angles helps to average over PSF residuals. %; thus the observing strategy depicted in Figure \ref{fig:os6} is appropriate regardless of whether RDI or ADI+RDI is ultimately used for PSF synthesis. 
RDI is preferred over ADI in situations where self-subtraction is significant: when there is little sky rotation or when targeting extended emission in circumstellar disks.

RDI requires PSF stability between the science and reference stars.
Therefore, it is commonly used in space-based observations\cite{Choquet2015} and will be the default observing strategy for CGI (Figure \ref{fig:os6}). 
Observations of bright reference stars are already required for CGI wavefront control, as described in Section \ref{sec:wavefront control}, and so the additional images needed for RDI can be added with minimal overhead.
RDI is less commonly used by ground-based facilities, where changing seeing conditions introduce PSF variability, and where target and reference star brightness must be closely matched in order to achieve similar AO correction quality. 
Nevertheless, ground-based RDI has been successfully used to improve sensitivity at small working angles in ``snapshot'' surveys where each target has little sky rotation\cite{Xuan2018} and in studies of disks\cite{Mawet2009, Rameau2012}.

Several factors must be considered when choosing a reference star. 
The star should not have any close companions or a circumstellar disk.
The stars should be close enough on the sky (typically $<20^\circ$ for CGI) that the telescope and instrument conditions do not change appreciably between reference and science pointings.
Both imaging and IFS modes are insensitive to stellar spectral type mismatch; the chromatic and time variations of the speckles dominate over object color differences.

\subsection{Spectral Differential Imaging}
SDI relies on the deterministic evolution of PSF speckles with wavelength.
When speckles are due only to phase aberrations in the wavefront, their location varies radially as $\lambda$, while their flux varies as the stellar spectrum modulated by $\lambda^{-2}$. 
Planets, on the other hand, remain at a fixed location and can have distinct spectral features (e.g.: CH$_4$ absorption).
These relationships can be used to synthesize the PSF at one wavelength from the PSF at another wavelength.
In its simplest form, SDI consists of subtraction of scaled images from two adjacent filters (e.g.: in and out of a CH$_4$ or $H\alpha$ features\cite{Racine1999, Marois2000a, Artigau2008, Dohlen2008, Close2014b}). 
The IFS implementation can take a more sophisticated approach, utilizing information from multiple spectral channels, but following the same underlying methodology\cite{Sparks2002}.
Current ground-based instruments operate in a regime where phase aberrations dominate over amplitude aberrations.

However, when speckles are induced by both phase and amplitude aberrations, the same wavelength scaling does not apply, and SDI breaks down. 
When phase and amplitude effects are comparable, as in CGI, the resulting speckle field is chromatic (Figure \ref{fig:psfs}, right panel). 
CGI cannot use SDI unless future algorithms are developed that can disentangle the contributions from the two types of speckles. 
Even though SDI cannot be used for PSF synthesis, the spectral information still adds value. 
Matched filtering, which compares the spectrum of a candidate to that of a planet model\cite{Ruffio2017}, can still be used to differentiate planets and residual speckles.

\subsection{Polarimetric Differential Imaging}
PDI\cite{Langlois2014, Perrin2015} can be used to extract polarized signals from companions or disks, under the assumption of polarization-independent speckles.
In this method, the target is observed at multiple (linear) polarization angles, and the polarization-invariant signal (the unpolarized stellar speckle field and the unpolarized astrophysical signal) is removed.
Unlike ADI, PDI has the benefit of preserving extended structures, so long as these structures are polarized.
PDI has enabled the ground-based characterization of a number of circumstellar disks and has achieved working angles smaller than those possible with ADI\cite{Perrin2015, 2017Msngr.169...32G}.

\begin{figure} [ht]
   \begin{center}
   \begin{tabular}{c} %% tabular useful for creating an array of images 
   \includegraphics[width=\textwidth]{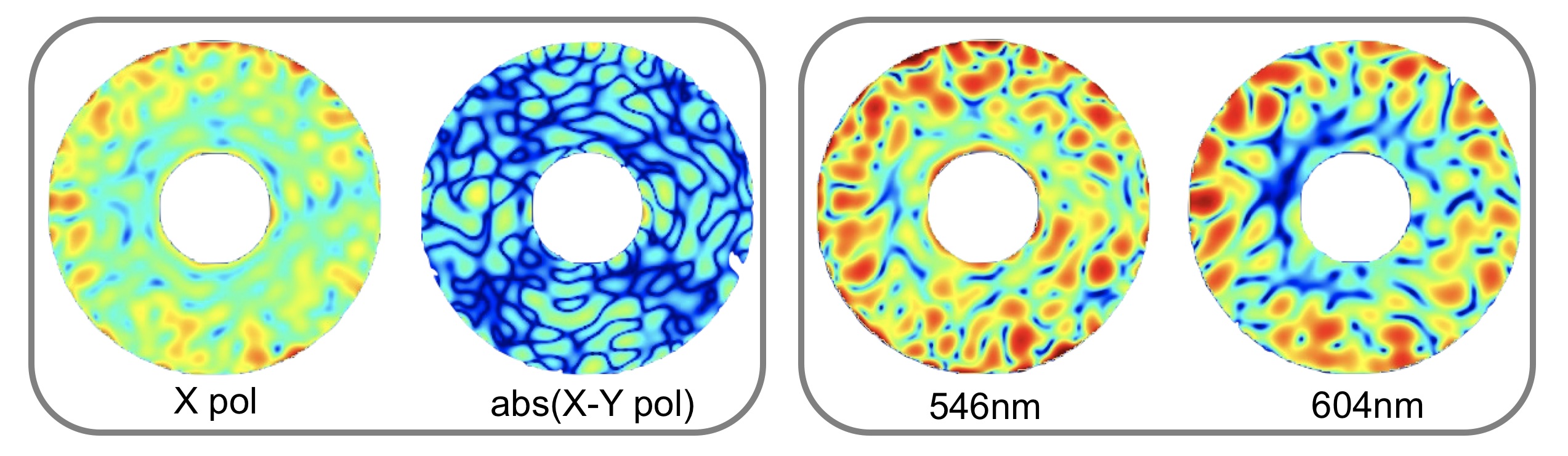}
   \end{tabular}
   \end{center}
   \caption[example]{ \label{fig:psfs} Simulated CGI speckle fields for imaging mode with the Hybrid Lyot Coronagraph, all with the same log scale display. Speckle fields vary with wavelength and are polarization-dependent.
\textit{Left:} Raw image in X polarization and the absolute difference of X and Y polarization images.
\textit{Right:} Raw images in the same polarization at 546nm and 604nm. Please refer to companion proceedings for additional information about CGI modeling efforts\cite{Krist2018,Rizzo2018,Zhou2018}.}
\end{figure}

In ground-based facilities, speckle fields are dominated by polarization-independent effects, but again, this is not the case for CGI. 
The fast telescope primary mirror and non-optimal mirror coatings induce polarization-dependent speckles, or ``polarization aberrations''\cite{Breckinridge2015, Krist2016, Krist2018} (Figure \ref{fig:psfs}, left panel). 
Future trade studies will investigate the possibility of observing in only a single polarization for improved dark hole optimization at the expense of throughput. 
Note that although PDI is not expected to improve CGI post-processed sensitivity, CGI will still be capable of polarimetry of bright sources (Section \ref{sec:pol cal}).

\section{CALIBRATION}

\subsection{Polarimetry}
\label{sec:pol cal}

Although CGI will not be capable of PDI, it will still be capable of (linear) polarimetry on bright sources with fluxes significantly above the polarization-dependent speckle noise floor. 
CGI is required to measure linear polarization fraction to better than 3\% systematic accuracy in the high signal to noise limit. 
CGI will have a set of four interchangeable linear analyzers (0$^\circ$, 45$^\circ$, 90$^\circ$, 135$^\circ$). 
Because CGI will not have a modulator, it will not be possible to self-calibrate the instrumental polarization from the science sequences, as is common practice in ground-based instruments with modulators\cite{Millar-Blanchaer2016, vanHolstein2017, Murakawa2004}. 

\begin{figure} [ht]
   \begin{center}
   \begin{tabular}{c}  
   \includegraphics[width=.6\textwidth]{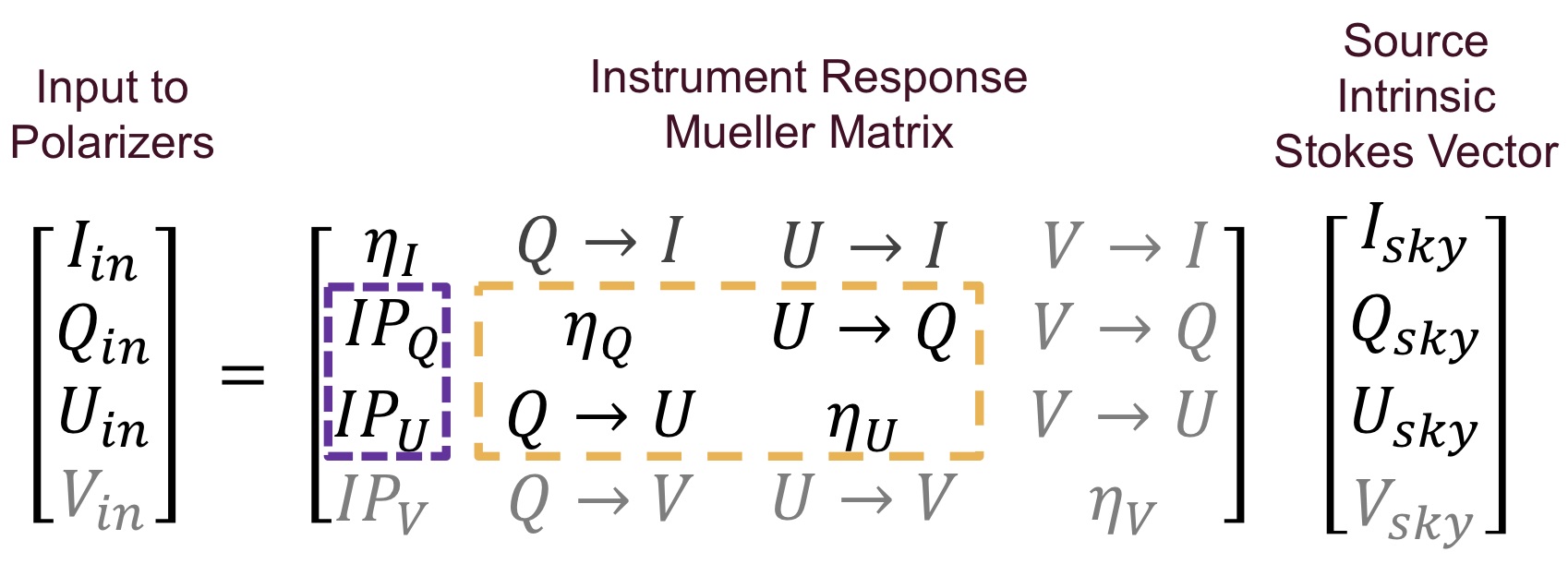}
   \end{tabular}
   \end{center}
   \caption[]{ \label{fig:pol cal} Effect of instrument response (Mueller Matrix) on the source intrinsic polarization. The quantities in the purple box should be calibrated on each target, while the quantities in the orange box should require less frequent calibration. CGI will not measure circular polarization and source-intrinsic circular polarization is expected to be small in most cases.}
\end{figure}

Routine observations of calibrators and telescope and instrument modeling will be critical. 
Figure \ref{fig:pol cal} shows the effect of throughput, crosstalk, and instrumental polarization on the astrophysical polarization signal. 
The instrumental polarization ($IP$) will have both constant terms and terms that vary with each target (i.e.: polarization-dependent speckles). 
Therefore, $IP$ should be calibrated on each science target, perhaps even after each optimization of the dark hole during a science sequence.
Efficiency in each polarization ($\eta$) and crosstalk between polarizations (eg: $Q\rightarrow U$) are likely to evolve more slowly, requiring less frequent calibration on polarized standard stars (cadence to be determined).
Catalogs of unpolarized and polarized standard stars\footnote{eg: \url{http://www.ukirt.hawaii.edu/instruments/irpol/irpol_stds.html}} may be supplemented with observations of pre-determined CGI PSF reference stars.
Note that CGI will not measure circular polarization (Stokes V), nor are circumstellar disks expected to be significantly circularly polarized. 
The effect of all crosstalk, including Stokes V, will have to be modeled to determine polarimetric efficiency and accuracy; this is the subject of ongoing efforts\cite{Breckinridge2015, Krist2017, Schmid2018, vanHolstein2018}.

\subsection{Astrometry and Photometry}

The high dynamic range and small field of view (FOV) of CGI images pose a significant challenge for both relative and absolute astrometric and photometric calibration. 
CGI will achieve single-frame flux ratios of $10^{-8}$ or better; this dynamic range exceeds detector capabilities, and so all science frames will be taken with the primary star occulted by the coronagraph. 
Without coronagraph masks, the FOVs of the imaging and IFS channels are 4.5'' and 1.1'' in radius, respectively; with coronagraph masks and corresponding field stops, the FOVs correspond to the detection limit curves shown in Figure \ref{fig:theplot}.
There are three primary challenges: determining the location of the star behind the coronagraph mask, flux calibrating coronagraphic images, and determining the plate scale, rotation, and distortion map. 
The latter requires periodic calibration, while the former two must be done for every science image and sequence. 

\begin{figure}[ht]
   \begin{center}
   \begin{tabular}{c}  
   \includegraphics[width=.4\textwidth]{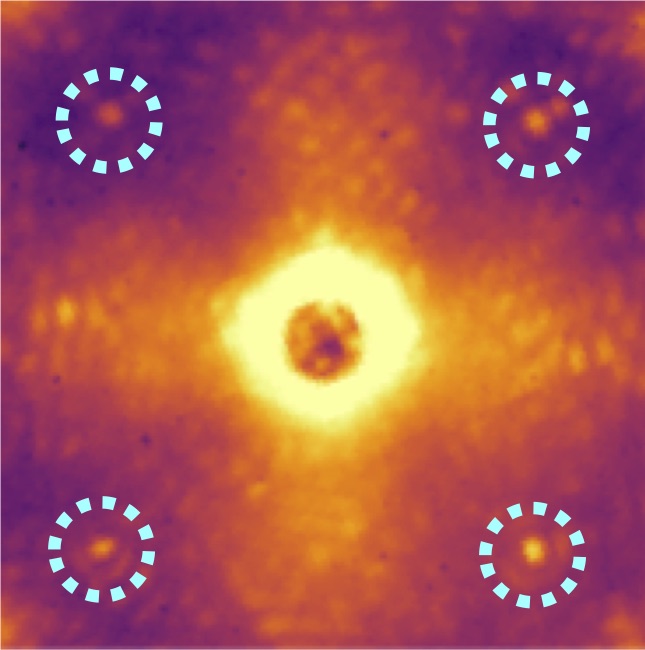}
   \end{tabular}
   \end{center}
   \caption[]{ \label{fig:sat spots} 
   A coronagraphic image with ``satellite spots'' (enclosed by blue dashed circles). Satellite spots provide both relative astrometry and photometry in each frame. CGI will inject satellite spots by placing one or more sinewaves on its deformable mirrors.}
\end{figure}

Many ground-based systems use ``satellite spots'' for relative astrometric and photometric calibration (i.e.: relative to the host star flux and location). 
Satellite spots are copies of the primary star PSF, injected at known flux levels and known offsets from the central star. 
There are two ways to inject satellite spots: pupil plane amplitude modulation (grids on pupil plane optics\cite{Sivaramakrishnan2010a}) or pupil plane phase modulation (sinewaves on deformable mirrors\cite{Jovanovic2015}). 
The former has the advantage of stability, but is also therefore inflexible; the location and amplitude cannot be adjusted. 
Satellite spots can also suffer from interactions with residual stellar speckles; the four spots in the GPI image in Figure \ref{fig:sat spots} show slightly different morphologies. 
If deformable mirror sinewaves are used, the amplitude and location of the spots can be adjusted on each target. 
Furthermore, the phase of the sinewaves can be modulated to average over quasistatic speckle interactions\cite{Jovanovic2015}. SCExAO uses this approach, and CGI would likely follow suit. 
One drawback of the sinewave method is that flux calibration relies on precise knowledge of the amplitude of the applied sinewaves; actuator calibration errors that would be removed in a closed-loop operation are not removed in this open-loop operation. 
DM calibration for CGI is the subject of ongoing efforts; the actuator accuracy will determine CGI's flux calibration accuracy.

Absolute photometric calibration of the satellite spots can be achieved in one of two ways: observations of an astrophysical binary system or self-calibration with a series of tiered satellite spots. 
The former requires a well-calibrated astrophysical point source with projected separation of several tenths of an arcsecond and a flux $\lesssim 100$ times the flux of the satellite spots ($10^{-5}-10^{-6}$). 
Such an object is beyond reach of all current visible-light high-contrast instruments, and so this method may be infeasible for CGI.
An alternative approach would employ a series images with satellite spots of varying flux ratios. 
A detailed CGI calibration plan is under development and will also investigate whether neutral density filters and/or variable detector gains are necessary.

Absolute astrometric calibration requires both an initial calibration, including a distortion map, and periodic checks of plate scale and orientation\cite{Maire2016,Wang2016a}. 
The initial calibration will be conducted pre-launch with calibration light sources. 
In-flight checks of the orientation and separations of the satellite spots could use relatively bright astrophysical calibrators (such as globular cluster fields), as the fluxes of the satellite spots could be significantly increased relative to their fluxes in science frames.
The astrometric fields would be calibrated by instruments on well-characterized observatories such as HST and Keck.
CGI could also be tied to the WFIRST Wide Field Instrument (WFI); a WFI snapshot taken at each pointing could be used for absolute astrometry, provided that the WFI-to-CGI mapping is calibrated periodically.

\section{INTEGRAL FIELD SPECTROGRAPH DATA PROCESSING}
The CGI IFS will follow in the footsteps of several high-contrast ground-based IFSs, including OSIRIS\cite{Larkin2006}, P1640\cite{Hinkley2011}, SPHERE\cite{Claudi2008}, GPI\cite{Larkin2014}, and CHARIS\cite{Peters2012}.
IFSs optically subdivide the focal plane into spatial pixels (``spaxels'') with mirrors or lenslets before passing the light through a disperser. 
The resulting array of ``microspectra'' (one low resolution spectrum per spaxel) are imaged onto the detector.
The CGI IFS will use a lenslet array and an $R\sim50$ prism, with a maximum bandpass of 20\%. The current baseline includes two 18\% bandpass filters that will span roughly 600--800~nm.

\begin{figure}[ht]
   \begin{center}
   \includegraphics[width=\textwidth]{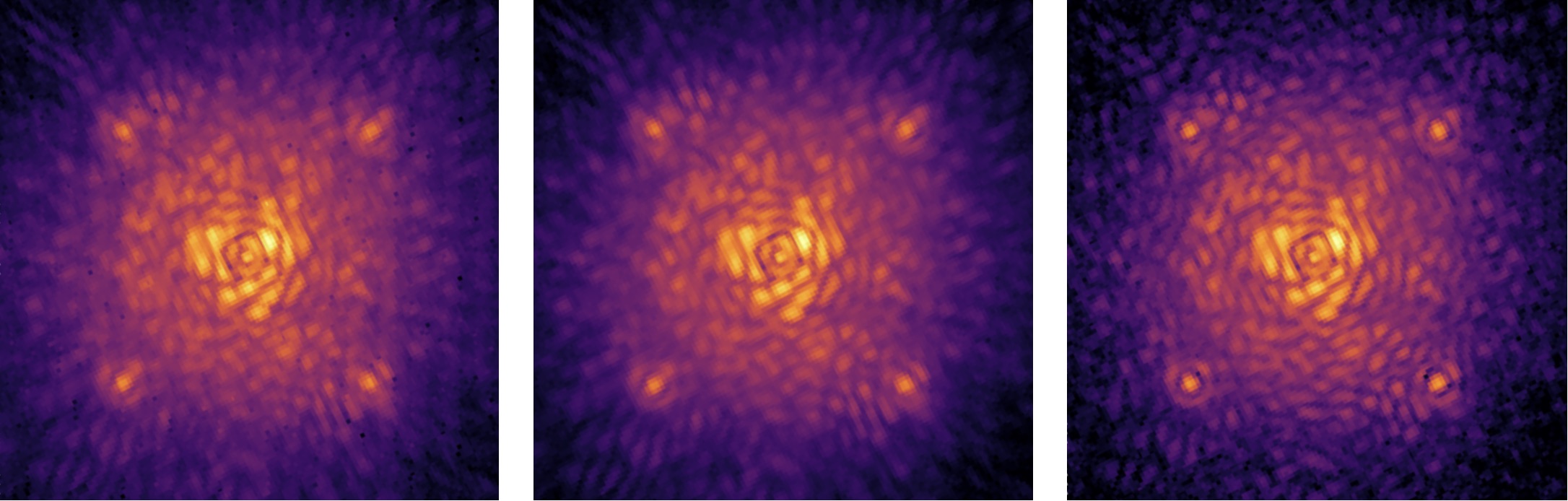}
   \end{center}
   \caption[]{ \label{fig:CHARIS redux} 
   Single wavelength channel of CHARIS broadband IFS data, reduced with three different spectral extraction techniques. From left to right: boxcar, Horne, and $\chi^2$. Image stretch is logarithmic and is the same for all three images.}
\end{figure}

Post-processing is required to reconstruct the 3D data cube (sky x, sky y, and wavelength) from the array of microspectra.
The most straightforward approach is boxcar (aperture) extraction; all pixels in the extraction region are given equal weight\cite{Mesa2015, Wang2017}.
A more sophisticated approach, Horne extraction, uses a wavelength-dependent axi-symmetric Gaussian PSF profile\cite{Horne1986}\footnote{Horne originally referred to this as ``optimal extraction,'' and this terminology is sometimes still used, although this method is not optimal in all cases.}. 
These approaches do not inherently account for wavelength channel covariance caused by the extended PSF wings\cite{Greco2016}.
A further refinement, least squares ($\chi^2$), uses high-fidelity lenslet PSFs to fit the spectrum with a linear combination of tophats of varying intensity and central wavelength, convolved with the lenslet PSFs\cite{Brandt2017}. 
This approach decreases the effects of channel-to-channel covariance and lenselet PSF variance across the FOV.
Figure \ref{fig:CHARIS redux} shows a single wavelength channel of CHARIS IFS data, extracted using each of these different techniques: boxcar, Horne, and $\chi^2$. 

Wavelength calibration requires knowledge of the detector x/y location and dispersion profile of each microspectrum.
For both ground-based instruments and CGI, calibration proceeds in two stages\cite{Wolff2016, Brandt2017, Zimmerman2011}. 
The initial calibration of lenslet locations and dispersion profiles is conducted off-sky/pre-launch with bright lab sources such as arclamps, tunable filters, or lasers.
The lenslet arrays may shift globally over the course of operations, and so the global translation and low-order distortion should be checked frequently, with cadence determined by the stability of the instrument. 
Ground-based instruments have the option to take snapshots of arclamps or other iternal sources for this purpose.
However, the CGI baseline design does not include an internal light source, and so wavelength calibration must be conducted on-sky; both 1\% narrowband observations of reference stars and observations of absorption or emission line targets are under consideration.

\section{SUMMARY}

The Coronagraph Instrument on the Wide Field Infrared Survey Telescope will demonstrate visible-light high-contrast imaging, polarimetry, and integral field spectroscopy at unprecedented sensitivity. 
CGI will enable the study of exoplanets and circumstellar disks in reflected visible light, at fluxes as low as $10^{-9}$ the flux of the primary star. 
This will be an improvement of roughly three orders of magnitude beyond current ground-based capabilities, and will constitute a shift from primarily near infrared wavelengths to wavelengths as low as 550~nm.

CGI can apply many of the lessons learned from ground-based high-contrast instruments in the areas of wavefront control, observing strategy, calibration, and integral field spectrograph data processing. 
Continuous control of low-order aberrations with a dedicated LOWFS/C system is needed for optimal sensitivity at small working angles. 
Sensing of non-common path low order aberrations and of high order aberrations is best achieved with focal plane WFS using science camera images themselves. 
Successful WFS/C requires precise knowledge of the instrument model and deformable mirror calibration. 
Spectral and polarimetric differential imaging are not likely to be applicable due to chromatic and polarization-dependent speckle behavior, and so CGI observing strategies will be designed around angular and reference differential imaging. 
Relative astrometric calibration will use ``satellite spots'' injected by sinewaves on the deformable mirror; absolute astrometry will then be calibrated against known systems and perhaps against the WFIRST Wide Field Instrument.
Relative photometry will also use satellite spots; absolute calibration of the flux of the spots requires precise knowledge of the deformable mirror shape.
A combination of on-sky calibrators and instrument models must account for instrumental polarization as well as crosstalk between polarization states and transmission efficiency of each linear state.
Spectral extraction requires both wavelength calibration and data processing techniques that account for covariances between wavelength channels.

\section*{ACKNOWLEDGEMENTS}       
 
\copyright 2018. 
This research was carried out in part at at the Jet Propulsion Laboratory, California Institute of Technology, under a contract with the National Aeronautics and Space Administration; government sponsorship acknowledged.

% References
\bibliography{report} % bibliography data in report.bib
\bibliographystyle{spiebib} % makes bibtex use spiebib.bst

\end{document}